Extreme Measures of Agricultural Financial Risk

By

John Cotter♣, Kevin Dowd* and Wyn Morgan♦

Abstract

Risk is an inherent feature of agricultural production and marketing and accurate measurement of it helps inform more efficient use of resources. This paper examines three tail quantile-based risk measures applied to the estimation of extreme agricultural financial risk for corn and soybean production in the US: Value at Risk (VaR), Expected Shortfall (ES) and Spectral Risk Measures (SRMs). We use Extreme Value Theory (EVT) to model the tail returns and present results for these three different risk measures using agricultural futures market data. We compare the estimated risk measures in terms of their size and precision, and find that they are all considerably higher than normal estimates; they are also quite uncertain, and become more uncertain as the risks involved become more extreme.

October 6, 2008

♣ Centre for Financial Markets, Smurfit School of Business, University College Dublin, Carysfort Avenue, Blackrock, Co. Dublin, Ireland. Email: john.cotter@ucd.ie. Cotter's contribution to the study has been supported by a University College Dublin School of Business research grant. Email:john.cotter@ucd.ie.

* Centre for Risk and Insurance Studies, Nottingham University Business School, Jubilee Campus, Nottingham NG8 1BB, United Kingdom. Email: Kevin.Dowd@nottingham.ac.uk.

♦ School of Economics, University of Nottingham, University Park, Nottingham NG7 2RD, United Kingdom. Email: wyn.morgan@nottingham.ac.uk.. (Corresponding author)

# 1. INTRODUCTION

The inherent variability in agricultural production (weather, pests, animal illness and so forth) alongside demand variations (food scares, fads, etc.) make for a marketing environment for farmers that is characterised by significant levels of risk (Moschini and Hennessy, 2001, Chern and Ricketsen (2003) and Carter and Smith (2007)). A natural question then arises - how do you measure the magnitude of risk being faced by agents? – and the last decade and a half have witnessed an explosion of research on different measures of financial risk, and especially on one particular measure, the Value-at-Risk (VaR). This 'VaR revolution' began when JP Morgan published its famous RiskMetrics model on the web in October 1994. VaR models were first used by financial institutions for their own risk management purposes, but have since been adopted by many non-financial corporates as well. Amongst their many uses, VaR models can be used to determine capital and reserving requirements, establish position limits and assess hedging strategies. They can also be used to manage cashflow, liquidity and credit risks as well as the market risks for which they were first developed. Estimation methods have improved considerably over the years, and the properties – and especially the limitations – of the VaR itself have become better understood. Various new measures of financial risk have also been proposed and these include, most notably, the coherent risk measures proposed by Artzner *et alia* (1999). These risk measures have the highly desirable property of sub-additivity, which the VaR lacks.[1] Thus, not only have VaR estimation methods improved over time, but there have also been improvements in the financial risk measures themselves, of which the VaR is but one.

---

[1] Suppose we let *X* and *Y* represent any two portfolios and let $\rho(.)$ be a measure of risk over a given forecast horizon. The risk measure $\rho(.)$ is subadditive if it always satisfies the condition $\rho(X+Y) \leq \rho(X) + \rho(Y)$. Subadditivity reflects the idea that risks should not increase, and should typically decrease, when we put them together, i.e., it reflects the notion that risks should diversify. The coherent risk measures are always sub-additive by construction, because sub-additivity is one of the axioms of coherence, but the VaR is not coherent and the failure of VaR to be sub-additive leads to the VaR having some strange and undesirable properties as a risk measure. See Artzner *et al.* (1999, p. 217, Dowd (2005, pp. 31-32)

The relevance of these developments to agricultural financial risks is self-evident. Yet, ironically, to date they have had only a limited impact on the agricultural economics and finance literature. Some indication of the current state of the art in agricultural financial risk measurement can be obtained from Table 1. This lists the main points of 8 different studies on this subject. Most of these studies use multivariate parametric approaches to estimate VaR, and these are typically based on the assumption that underlying risks factors are multivariate normally distributed. Some studies also use historical simulation methods to estimate the VaR. One study (Zhang *et al.* (2007)) uses Monte Carlo methods, and two (Siaplay *et al.* (2005) and Odening and Hinrichs (2003)) include results based on Extreme-Value Theory (EVT). It is also noteworthy that all but one of these studies focuses exclusively on the VaR risk measure.[2] To our knowledge, there are no studies so far of coherent risk measures applied to agricultural risk problems.

**Insert Table 1 here**

This paper examines three different measures of financial risk applied to agricultural risk. The measures examined are the VaR and two members of the family of coherent risk measures. The first of the coherent risk measures is the Expected Shortfall (ES), which is loosely speaking the average of the 'tail losses' or losses exceeding the VaR. The ES takes account of the magnitude of losses exceeding the VaR. This, and the related fact that it is subadditive, makes the ES a superior risk measure to the VaR on *a priori* grounds. However, both the VaR and ES measures depend on the choice of a confidence level that delineates the cutoff to the tail region, and there is seldom an 'obvious' choice of what the confidence level should be. Moreover, the ES has the undesirable property of implying that the user is risk-neutral, and this sits uncomfortably with the use of such measures by risk-averse agents in the first place.[3]

---

[2] The one exception (Zhang *et al.*, 2007) looks at lower partial moment measures based on the downside risk literature (e.g., Fishburn, 1977) rather than the coherent risk measures that have been much discussed in the mainstream financial risk literature. The VaR and the ES can be regarded as special cases of the lower partial moment measures if the lower partial moment parameter takes the values 0 or 1 respectively (see Dowd, 2005, p. 26).

The other coherent risk measure is a Spectral Risk Measure (SRM) proposed by Acerbi (2002, 2004). The distinctive feature of an SRM is that it specifically incorporates a user's degree of risk aversion. Since SRMs are a subset of the family of coherent risk measures, they have the attractions of coherent risk measures as well. A tractable type of SRM is that based on an exponential risk aversion function, and a nice feature of exponential risk aversion function is that the extent of risk aversion depends on a single parameter, the coefficient of absolute risk aversion $R$. Once a user chooses the value of $R$ that reflects its attitude to risk, it can then obtain an 'optimal' risk measure that directly reflects its degree of risk aversion. So, whereas the VaR or ES are contingent on the choice of an arbitrary parameter, the confidence level, whose 'best' value cannot easily be determined, a spectral-coherent risk measure is contingent on a parameter whose 'best' value can be selected by the agent that uses it.

Our measurement of the three risk measures is for corn and soybean spot and futures contracts as these goods represent an important element of US agricultural production: corn due to its role in feed grain production and soybeans for vegetable oil production. We analyse the contracts for both long and short positions whose risk would be of interest to different possible users such as farmer producers and processors.

The focus of this study is extreme financial risk – the risk associated with the prospect of low probability, high impact losses. There has been considerable interest in extreme risks over the last decade. The literature on extremes tells us that extremes should be modelled separately from the rest of the distribution using the distributions implied by Extreme Value (EV) theory,[4] and should *not* be modelled by fitting full distributions to the data in an ad hoc way (e.g., such as assuming Gaussianity). In essence, it suggests that we can either model the extremes themselves using one of the Generalised EV distributions implied by the Extreme Value Theorem or we can model the exceedances over a high threshold using a Generalised Pareto Distribution (GPD; see, e.g., Embrechts et *alia* (1997)). This latter approach is often referred to as the Peaks-Over-Threshold approach. We choose the latter because it (typically) involves one less parameter and because it fits more easily with the likelihood that extreme losses occur in clusters. The application of the GPD can be justified by theory that tells us that the tail observations should follow a GPD in the asymptotic limit as the threshold gets bigger. Once the GDP curve is fitted to the data, it can then be extrapolated to give estimates of any extreme quantiles or tail probabilities we choose.

---

[3] For its part, the VaR is even worse, as it implies that a user who chooses to use the VaR as a risk measure must be highly risk-loving (see Cotter and Dowd, 2007, p. 3472).
[4] For more on EVT, see, e.g., Embrechts *et al.*,1997, or Beirlant *et al.*, 2004. Note that tail risk measures are underestimated using Gaussianity and this estimation bias deteriorates as one moves further out into the tail (Cotter, 2007).

Accordingly, in this paper, we use the POT approach to estimate and compare the extreme VaRs, ESs and SRMs for corn and soybean contracts. Bearing in mind that the usefulness of any estimates of financial risk measures also depends crucially on their precision, we also examine alternative methods of estimating their precision.[5] Given the heavy reliance of Gaussianity in the literature, we also produce estimates of risk measures using Gaussianity.

This paper is organised as follows. Section 2 reviews the risk measures to be examined. Section 3 reviews the Peaks-Over-Threshold (POT) approach and section 4 details the POT-based risk measures. Section 5 introduces the spot and futures corn and soybean data used in our empirical work and provides some preliminary data analysis. Section 6 describes the bootstrap procedure used to derive the precision metrics used in the paper. Section 7 then estimates VaR and ES, and section 8 estimates the SRMs. Each of these sections also examines the precision of these estimated risk measures. Section 9 concludes.

---

[5] As noted already in the text, two of the studies listed in Table 1 present results based on EVT. Of these, Siaplay *et alia* (2005) report EV estimates of VaR in a single table obtained using the EV function in Palisade Corporation's '@Risk' package, but provide no EV analysis as such. We also note there that this function only allows the user to model a Gumbel EV distribution, and this distribution is not compatible with heavy-tailed returns. Odening and Hinrichs (2002) provide an analysis based on Generalised EV theory, but they report rather unstable estimates of the tail index parameter – a common problem in this area - and this makes their results unreliable.

## 2. MEASURES OF FINANCIAL RISK

Suppose *X* is a realised random loss variable – a variable that assigns loss outcomes a positive sign and profit outcomes a negative one - for a commodity over a given horizon. If the confidence level is $\alpha$, the VaR at this confidence level is:

$$VaR_\alpha = q_\alpha \tag{1}$$

where the term $q_\alpha$ is the $\alpha$-quantile of the loss distribution. For any given horizon, the VaR is defined in terms of its conditioning parameter, the confidence level, which is arbitrarily specified by the user. Viewed as a function of the quantiles of the loss distribution, it is useful to note here that the VaR places all its weight on a single quantile that corresponds to the chosen confidence level and places no weight on any others. This implies that the user only 'cares' about a single loss quantile, and is not concerned about higher losses, and it is this rather strange property that causes the VaR risk measure to be non-subadditive (Acerbi, 2004).

The second measure, the ES, gives equal weight to each of the worst $1-\alpha$ of losses and no weight to any other observations. The ES is superior to the VaR in a number of respects (e.g., it is subadditive and coherent and because takes account of losses beyond the VaR quantile). However, the ES is specified in terms of the same conditioning parameter as the VaR and, as with the VaR, there is generally little to tell us what value this parameter should take.

Our third measure is the Spectral Risk Measure (SRM). Following Acerbi (2002), consider a risk measure $M_\phi$ defined by:

$$M_\phi = \int_0^1 q_p \phi(p) dp \tag{3}$$

where $q_p$ is the $p$ loss quantile, $\phi(p)$ is a weighting function defined over $p$, the cumulative probabilities in the range between 0 and 1. Borrowing from Acerbi (2004, proposition 3.4), the risk measure $M_\phi$ is coherent if and only if $\phi(p)$ satisfies the following properties:

- *Positivity*: $\phi(p) \geq 0$, i.e., weights are always non-negative.
- *Normalisation*: $\int_0^1 \phi(p)dp = 1$, i.e., weights sum to one.
- *Increasingness*: $\phi'(p) \geq 0$, i.e., higher losses have weights that are higher than or equal to those of smaller losses.

We now need to specify a suitable weighting (or risk-aversion) function and a reasonable choice is the exponential risk-aversion function:

$$\varphi(p) = \frac{R e^{-R(1-p)}}{1 - e^{-R}} \qquad (4)$$

where $R>0$ is the coefficient of absolute risk aversion. This weighting/risk-aversion attaches higher weights to larger losses, and, moreover, the weights rise more rapidly as the user becomes more risk-averse.

The value of the risk measure can then be obtained by substituting (4) into (3), viz.:

$$M_\varphi = \int_0^1 \frac{R e^{-R(1-p)}}{1 - e^{-R}} q_p \, dp = \frac{R e^{-R}}{1 - e^{-R}} \int_0^1 e^{Rp} q_p \, dp \qquad (5)$$

# 3. THE PEAKS OVER THRESHOLD (GENERALISED PARETO) APPROACH

We model the agricultural tail risks using a Peaks over Threshold (POT) approach which focuses on the realisations of a random variable $X$ over a high tail threshold $u$. More particularly, if $X$ has the distribution function $F(x)$, we are interested in the distribution function $F_u(x)$ of exceedances of $X$ over a high tail threshold $u$:

$$F_u(x) = P\{X - u \leq x | X > u\} = \frac{F(x+u) - F(u)}{1 - F(u)} \qquad (6)$$

As $u$ gets large, the distribution of exceedences tends to a Generalized Pareto Distribution (GPD):

$$G_{\xi,\beta}(x) = \begin{cases} 1 - (1 + \xi x/\beta)^{-1/\xi} & \text{if } \xi \geq 0 \\ 1 - \exp(-x/\beta) & \xi < 0 \end{cases} \qquad (7)$$

where

$$x \in \begin{cases} [0, \infty) \\ [0, -\beta/\xi] \end{cases} \text{ if } \begin{array}{l} \xi \geq 0 \\ \xi < 0 \end{array}$$

and the shape $\xi$ and scale $\beta > 0$ parameters are estimated conditional on the threshold $u$ (Balkema and de Haan, 1974; Embrechts *et al.*, 1997, pp. 162-164). *En passant*, note that the shape parameter $\xi$ sometimes appears in GPD discussions couched in terms of its inverse, a tail index parameter $\alpha$ given by $\alpha = 1/\xi$.

The behavior of the GPD tail depends on the values of these parameters, and the shape parameter is especially important. A negative $\xi$ is associated with very thin-tailed distributions that are rarely of relevance to financial data, and a zero $\xi$ is associated with thin tailed distributions such as the Gaussian, but the most relevant for our purposes are heavy-tailed distributions associated with $\xi > 0$. The tails of such distributions decay slowly and follow a heavy tailed 'power law' function. Moreover the number of finite moments is determined by the value of $\xi$ (or $\alpha$): if $\xi \leq 0.5$ (or, equivalently, $\alpha \geq 2$), we have infinite second and higher moments; if $\xi \leq 0.25$ (or $\alpha \geq 4$), we have infinite fourth and higher moments, and so forth. $\alpha$ therefore indicates the number of finite moments. Evidence generally suggests that the second moment is probably finite, but the fourth moment is more problematic (see, e.g., Loretan and Phillips,1994).

The values of the GPD parameters can be estimated by Maximum Likelihood (ML) methods using suitable (e.g., numerical optimization) methods. The log-likelihood function of the GPD is:

$$l(\xi,\beta) = -n(\ln(\beta) - (1+1/\xi)\sum_{i=1}^{n}\ln(1+\xi x_i/\beta) \qquad \text{for } \xi \neq 0 \qquad (8)$$

$$l(\beta) = -n(\ln(\beta) - \beta^{-1}\sum_{i=1}^{n}x_i \qquad \text{for } \xi = 0 \qquad (9)$$

where in both cases $x_i$ satisfies the constraints specified above for $x$.

## 4. FORMULAS FOR RISK MEASURES UNDER THE POT APPROACH

Assuming that $u$ is sufficiently high, the distribution function for exceedances is given by:

$$F_u(x) = 1 - \frac{N_u}{n}\left(1+\xi\frac{x-u}{\beta}\right)^{\frac{-1}{\xi}} \qquad (10)$$

where $n$ is the sample size and $N_u$ is the number of observations in excess of the threshold (Embrechts *et al.*,1997, p. 354). The $p^{th}$ quantile of the return distribution - which is also the VaR at the (high) confidence level $p$ – can then be obtained by inverting the distribution function, viz.:

$$q_p = VaR_p = u + \frac{\beta}{\xi}\left\{\left(\frac{n}{N_u}p\right)^{-\xi} - 1\right\} \tag{11}$$

The ES is then given by:

$$ES_p = \frac{q_p}{1-\xi} + \frac{\beta - \xi u}{1-\xi} \tag{12}$$

To obtain our SRM, we now substitute (11) into (5) to get:

$$M_\phi = \int_0^1 \phi(p) q_p(X) dp = \int_0^1 \frac{e^{-(1-p)/R}}{R(1-e^{-1/R})}\left[u + \frac{\beta}{\xi}\left\{\left(\frac{n}{N_u}p\right)^{-\xi} - 1\right\}\right] dp \tag{13}$$

Having obtained the risk-measure formulas, estimates of the risk measures themselves are then obtained by estimating/choosing the relevant parameters and plugging these into the appropriate (i.e., (11) for the VaR, (12) for the ES, and (13) for the SRM). This is straightforward for the VaR and the ES; however, for spectral risk measures, we need to use a suitable numerical integration method (e.g., a trapezoidal rule, Simpson's rule, etc.: see Miranda and Fackler, 2002, or Cotter and Dowd, 2006, for further details).

## 5. DATA AND PRELIMINARY ANALYSIS

Our data set consists of weekly logarithmic price changes for Corn and Soybean contracts traded on the CBOT between January 1979 and December 2006 totalling 1461 observations. For each product there are 8 series analysed: 1 futures and 7 spot across 7 different geographical areas. We examine the tails of both long and short positions for each series, thus giving us a total of $8\times 2\times 2=32$ cases in total. We choose these particular crops for their importance in the US agricultural sector. Corn is the most widely produced feed grain in the US and accounts for 90% of the total value of feed grains produced. Approximately 80 million acres are planted to corn with most being in the heartland states. Illinois is the largest producer along with Iowa, hence the focus on the former for this analysis. Soybeans are also selected as the US is the world's largest producer and exporter of them and approximately 2.5 billion bushels were produced in 2007[6]. Illinois is again a major producer and is second only to Iowa in output terms. Soybeans are used for vegetable oil production and the meal for animal feed. Thus, we believe our choice of crop and state captures significant agricultural activity and thus could be viewed as suitably representative of arable production in the US albeit with a constrained focus.

As a preliminary, we illustrate some indicative time series properties in Figure 1 and Table 2. The mean returns are near zero for both spot and futures contacts, and the corresponding standard deviations suggest weekly volatilities in excess of 3% for both sets of contracts. The series are mostly negative skewed and always have excess kurtosis, and Jarque-Bera results indicate that normality is always rejected.

**Insert Figure 1 here**
**Insert Table 2 here**

Despite the fact that normality is rejected so strongly, it is useful to know what the risk measures would be under the counterfactual and heavily used assumption that returns are normal. These are reported in Table 3, and we will comment on these later when presenting the POT estimates of these risk measures.

**Insert Table 3**

---

[6] Data are drawn from the National Agricultural Statistics Service (NASS) of the USDA website at http://www.nass.usda.gov/

Figure 2 shows QQ plots for these series' empirical return distributions relative to a normal (or Gaussian) distribution. If the normal distribution is an adequate fit, then the QQ plot should be approximately a straight line. However, in each case, we find that the QQ plot is approximately straight only in the central region, and that the tails show steeper slopes than the central observations: this indicates that the tails exhibit heavier kurtosis than the normal distribution, and is consistent with the results of Table 2.

**Insert Figure 2 here**

In addition, the points where the QQ plots change shape provides us with natural estimates of tail thresholds, and these implied thresholds are also consistent with the tail index plots – plots of the estimated tail index $\alpha$ and its 95% confidence interval against the number of exceedances – shown in Figure 3. The number of exceedances reflects the choice of threshold, a smaller number reflecting a higher threshold. In each case the estimated tail index is stable over a wide range of exceedance numbers (or threshold size, if you prefer), and this tells us that the estimated indices are stable relative to the thresholds selected.

**Insert Figure 3 here**

The approach taken here focuses on both short and long positions. The rationale for this is to reflect the various agents that operate in the supply chain for agricultural commodities. At one end, the farmer faces price risk through production and thus will be interested in short positions. Equally, processors (and possibly retailers if the product can be sold without much processing such as potatoes for example) are concerned about the price of their inputs rising and will tend to take long positions in the futures market. Finally, there are also merchants who both buy and sell the commodities and potentially face input and output price risk and thus could take a mixed strategy approach to trading by going both short and long depending on their circumstances.

We now fit the distributions of exceedances and ML estimates of the GPD parameters are given in Table 4 for both long and short trading positions. The Table gives the assumed thresholds *u*, the associated numbers of exceedances ($N_u$) and the observed exceedance probabilities (*prob*). Also included and of most interest for the risk measures are the tail indices, $\xi$, and the scale parameter, $\beta$. The tail indices are generally positive (though not statistically significant) for the spot and futures contracts, and the scale parameters vary around 2. The numbers and probabilities of exceedances vary somewhat, but all confirm that the chosen thresholds are in the stable tail-index regions identified earlier.

**Insert Table 4 here**

To check that the GPD provides an adequate fit, Figure 4 shows empirical exceedances fitted to the GPDs based on the parameter estimates given in Table 1, and the results confirm that the GPD provides a good fit in all cases.

**Insert Figure 4 here**

## 6. BOOTSTRAP ALGORITHM

The estimates of standard errors and confidence intervals reported in this paper were obtained using a semi-parametric bootstrap set out by Cotter and Dowd (2006). To implement this procedure, we begin by taking 5000 bootstrap resamples, each of which consists of *n*=1492 uniform random variables. Each resample is then sorted into ascending order so that its relative frequencies can be considered 'as if' they were a set of resampled cumulative probabilities. For example, for the $j^{th}$ resample, these relative frequencies are as $p_1^j, p_2^j, ..., p_n^j$, where $p_i^j \leq p_{i+1}^j$. We then use the fitted GPD (i.e., (11)) to obtain each element of the $j^{th}$ resample set of losses. Thus, if $p_i^j$ is the $i^{th}$ cumulative probability in the $j^{th}$ resample, then $q_i^j$, the $i^{th}$ highest loss in the $j^{th}$ resample, can be obtained from

$$q_i^j = \hat{u} + \frac{\hat{\beta}}{\hat{\xi}}\left\{\left(\frac{n}{\hat{N}_u}p_i^j\right)^{-\hat{\xi}} - 1\right\} \qquad (14)$$

where (14) is a version of (11) in which the '^' refer to the sample-based estimates of the GPD parameters. Since the VaRs are quantiles, (14) gives us direct resample estimates of the VaRs. Resample estimates of the ES and SRM are then obtained using (12) and (13) respectively (with $q_p$ replaced by $q_i^j$ and parameters replaced by their '^' estimates). For each resample, the standard errors and confidence interval were obtained from the set of resample estimates of the appropriate risk measures.

## 7. ESTIMATES OF VALUE AT RISK AND EXPECTED SHORTFALL

GPD estimates of VaR and ES are given in Table 5 for confidence levels of 99%, 99.5% and 99.9%: Table 5a gives the results for corn contracts and Table 5b gives the results for soybean contracts. To illustrate, the VaR of 9.989 at the 99% level implies that there is a 1% chance of having losses greater than 9.989% of the value of the corn Region 1 contract for a long trading position. These show, as we might expect, that estimated risk measures rise with the confidence level, and that the estimated VaRs are notably larger than the estimated ESs. There are no great differences between the different contracts or between the corn and soybean estimates of the risk measures, but the short and long results can be somewhat different from each other. It is also noteworthy that the estimated risk measures are usually much higher than the Gaussian-based estimates in Table 3 and the divergence increases as one moves to more extreme probability levels. This suggests that extreme risks are large, and that assuming Gaussianity in these circumstances can lead to very considerable underestimates of our risk measures.

The Table also reports the bootstrapped standard errors of the estimated risk measures, and these rise considerably with the confidence level: this indicates that estimated risk measures become considerably less precise as the confidence level rises. This is a well-known phenomenon, and reflects the fact that as the confidence level rises, we are dealing with an increasingly extreme tail measured with fewer and fewer observations.[7]

**Insert Table 5 here**

Table 6 shows bootstrapped estimates of the standardized 90% confidence intervals for the VaR and ES: these are estimates of the 90% confidence intervals divided by the estimated mean risk measure, and are easier to interpret than conventional confidence intervals. So, for example, the first two results in the first row of Table 6a tell us that the 90% confidence interval for the region 1 spot VaR varies from 89.3% to 111.7% of the mean VaR, and so forth. Two features of these results stand out:

- The standardized confidence intervals for the ES are generally a little narrower than those for the VaR: this confirms that in relative terms, estimates of the ES are more precise than estimates of the VaR.
- The confidence intervals are fairly symmetric for the risk measures predicated on the 99% confidence level, but become asymmetric as the confidence level rises and, in particular, we see that the right bound is further from the mean risk measure than the left bound. To give an example, at the 99.9% confidence level, the standardized confidence interval spans the range from 80% to 125.5% of the mean risk measure (i.e., down 20%, but up 25.5%). This finding is also to be expected and again reflects the fact that as we move further out into the extreme tail, we run into fewer observations and our uncertainty increases further.

---

[7] Interestingly, we also see that the standard errors are usually only a little larger for the ESs than for the VaRs: these indicate that ES estimates are a little less precise than VaR estimates in absolute terms. However, the ratios of estimated risk measures to standard errors are often lower for the ESs than for VaRs, so in relative terms (i.e., taking account of the sizes of the two risk measures), it is often the case that the ES is more precisely estimated than the VaR.

**Insert Table 6 here**

## 8. ESTIMATES OF SPECTRAL RISK MEASURES

We now turn to estimates of spectral risk measures. As we have discussed already, these risk measures make use of the coefficient of absolute risk aversion $R$ rather than the confidence level as their conditioning parameter. The value of this coefficient depends on the user's attitude to risk, and can in principle be any positive number (assuming that the user is in fact risk-averse). However, in the present EV context it only makes sense to work with fairly high values of $R$: the higher is $R$, the more we are concerned about very high (i.e., extreme) losses relative to more moderate ones. A concern with extremes therefore suggests a high value of $R$. Accordingly, we consider here values of $R$ equal to 20, 100 and 200.

Once a value of $R$ has been chosen, we can estimate the value of the integral (13) using numerical integration. The idea behind this is to discretize the continuous variable *p* into a large number *N* of discrete 'slices', where the discrete approximation gets better as *N* gets larger. We then choose a suitable numerical integration method, and the ones we considered were the trapezoidal rule, Simpson's rule, and numerical integration procedures using quasi-Monte Carlo methods based on Niederreiter and Weyl algorithms respectively.[8]

However, we first need to evaluate the accuracy of these methods. To help us do so, Table 7 gives estimates of the approximation errors generated by these alternative numerical integration methods based on alternative values of *N* and a plausible set of benchmark parameters.[9] These results indicate that all methods have a negative bias for relatively small values of *N*, but they also indicate that the bias disappears as *N* gets large. In addition, they suggest that for high *N*, the trapezoidal method is at least as accurate as any of the others.

---

[8] The choice of numerical integration method was also influenced by the need to have fast integration algorithms for use in our bootstrap algorithms. We used the Miranda-Fackler (2002) CompEcon functions, which are very fast indeed.

**Insert Table 7 here**

For the remaining estimations, we selected a benchmark method consisting of the trapezoidal rule calibrated with *N*=1 million.[10]

Estimates of SRMs and their bootstrapped standard errors and standardised 90% confidence bounds are given in Table 8. In many respects these results are comparable to those obtained earlier for the VaR and ES, but with $R$ playing the same role as the earlier confidence level. In particular, we see that:

- Estimated SRMs are considerably higher than the normal estimates in Table 3.
- Estimated SRMs rise notably as $R$ increases.
- Estimated SEs and the widths of confidence intervals rise as $R$ increases; we also see some asymmetry in the confidence intervals for very high values of $R$, again with the right bound being a little further away from the mean than the left bound.[11]
- Differences across contract types are fairly small, and the only noteworthy difference between the corn and soybean results is that the latter have more pronounced differences between long and short positions.

**Insert Table 8 here**

---

[9] These benchmark parameters were the mean parameters in Table 2 for the case where $R = 100$.

[10] There is of course a tradeoff between calculation time and accuracy, but the choice of *N*=1 million gives us results that are accurate to within half a percentage point in the illustrative case examined in Table 7, and this is accurate enough for our purposes.

[11] This phenomenon was also observed by Cotter and Dowd (2006), and the explanation is that as $R$ increases, an SRM estimator places more weight on a smaller number of extreme observations, and therefore operates with a smaller effective sample size. For very high values of $R$, we would then get right- and left-asymmetry reflecting the greater paucity of observations on the right-hand side.

## 9. CONCLUSIONS

Effective and accurate measurement of risk in agricultural markets is central to informing how best to design strategies and instruments aimed at helping farmers manage the risks they face. Toward this end, this paper applies the Peaks-Over-Threshold version of Extreme Value Theory to estimate the extreme financial risk measures for a selection of agricultural contracts. The risk measures considered were the Value at Risk (VaR), the Expected Shortfall (ES), and the Spectral Risk Measure (SRM) based on an exponential risk-aversion function for a given coefficient of absolute risk aversion. We examine the properties of these risk measures and suggest that SRM is to be preferred to the ES, which in turn is to be preferred to the VaR. We also estimate both the risk measures themselves and some precision metrics obtained using a parametric bootstrap procedure. Our empirical results suggest three main conclusions, and this is the case for all three risk measures. First, we find that the estimated risk measures are all considerably higher than the estimates we would have obtained under Gaussianity. This suggests that Gaussianity can lead to major under-estimates of extreme risks. Second, we find that estimated risk measures are quite uncertain, as judged by the estimated standard errors and confidence intervals. This is to be expected, as EV problems almost by definition involve small numbers of extreme observations. Third, we find that the degree of uncertainty associated with our estimated risk measures increases as we go further out into the tail. This finding also makes intuitive sense: the further we go into the tail, the more sparse our observations become, and the more uncertain any estimates will be. In a nutshell, extreme risk measures are large, but also uncertain.

**REFERENCES**

Acerbi, C., 2002. Spectral measures of risk: a coherent representation of subjective risk aversion. Journal of Banking and Finance 26: 1505-1518.

Acerbi, C., 2004. Coherent representations of subjective risk-aversion, in G. Szego (Ed), Risk Measures for the 21$^{st}$ Century, Wiley, New York, pp. 147-207.

Artzner, P., F. Delbaen, J.-M. Eber, and D. Heath, 1999. Coherent measures of risk. Mathematical Finance 9, 203-228.

Balkema, A. A., and L. de Haan, 1974. Residual lifetime at great age. Annals of Probability 2, 792–804.

Beirlant, J. , Y. Goegebeur, J. Segers and J. Teugels (2004) Statistics of Extremes: Theory and Applications. New York: Wiley.

Carter, C. and Smith, A. (2007) "Estimating the Market Effect of a Food Scare: The Case of Genetically Modified Starlink Corn", The Review of Economics and Statistics, 89(3), pp 522-533.

Chambers, R.G. and Quiggin, J. (2004) "Technological and Financial Approaches to Risk Management in Agriculture: an integrated approach", The Australian Journal of Agricultural and Resource Economics, 48(2) pp. 199–223

Chern C. and Rickertsen, K. (eds) (2003) Health, Nutrition and Food Demand, Wallingford: CABI

Cotter, J., and K. Dowd, (2006), Extreme Spectral Risk Measures: An Application to Futures Clearinghouse Margin Requirements, Journal of Banking and Finance, 30, 3469-3485.

Cotter, J., 2007. Varying the VaR for unconditional and conditional environments, Journal of International Money and Finance, 26, 1338-1354.

Dowd, K., 2005, Measuring Market Risk, Second edition, Chichester: John Wiley and Sons.

Dawson, P. J., and B. White, 2005. Measuring price risk on UK arable farms. Journal of Agricultural Economics, 56, 239-252.

Embrechts, P., Kluppelberg C., Mikosch, T., 1997. Modelling Extremal Events for Insurance and Finance. Springer Verlag, Berlin.

Fishburn, P. C., 1977. Mean-risk analysis with risk associated with below-target returns. American Economic Review 67, 116-126.

Loretan, M., and P. C. B. Phillips, 1994. Testing the covariance stationarity of heavy-tailed time series. Journal of Empirical Finance, 1, 211-248.

Manfredo, M.R. and R.M. Leuthold, 2001. Market risk and the cattle feeding margin: an application of Value at Risk. Agribusiness, 19, 333-353.

Miranda, M. J., and P. L. Fackler, 2002. Applied Computational Economics and Finance. MIT Press, Cambridge MA and London.

Miranda, Mario J. and Joseph W. Glauber, 1997 "Systemic risk, reinsurance, and the failure of crop insurance markets, " American Journal of Agricultural Economics, 79 (1) pp 206-215.

Moschini, G. and Hennessy, D.A. (2001) Uncertainty, risk aversion, and risk management for agricultural producers. Handbook of Agricultural Economics, Boston: Kluwer

Odening, M. and J. Hinrichs, 2003. Using extreme value theory to estimate value-at-risk. Agricultural Finance Review, Spring, pp 5-73.

Odening, M., and O. Mußhoff, 2002. Value-at-Risk – ein nützliches Instrument des Risikomanagements in Agrarbetrieben? In M. Brockmeier et alia (eds) Liberalisierung des Weltagrarhandels – Stratagien und Konsequenzen. Schriften der Gesellschaft für Wirtschafts- und Sozialwissenschaften des Landbaus, Band 37.

Pritchett, J. G., G. F. Patrick, K. J. Collins and A. Rios (2004) Risk management strategy evaluation for corn and soybean producers. Agricultural Finance Review, Spring, 45-60.

Siaplay, M., W. Nganje, and S. Kaitibie, 2005. Value-at-risk and food safety losses in turkey processing. Agribusiness and Applied Economics Report No. 57, Department of Agribusiness and Applied Economics, North Dakota State University.

Wilson, W.W., W. E. Nganje, and C. R. Hawes, 2005, Value at risk in bakery procurement. Review of Agricultural Economics 29, 581-595.

Zhang, R, J. E. Houston, D. V. Vedenov, and B. J. Barnett, 2007. "Hedging downside risk to farm income with futures and options: effects of government payment
19

programs and federal crop insurance plans" paper presented at the American Agricultural Economics Association annual meeting, Portland, Oregon, August 2007.

# FIGURES

## Figure 1a: Time Series Plots of Weekly Series: Corn

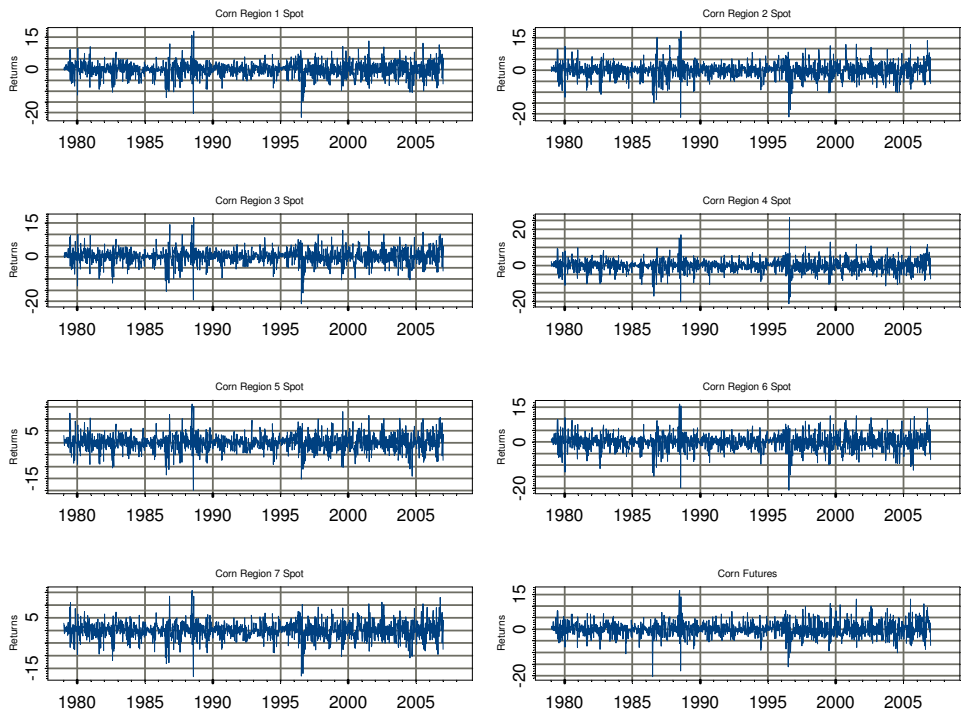

Notes: Plots show weekly % returns for each contract over the period January 1979 to December 2006. The sample size is 1461.

**Figure 1b: Time Series Plots of Weekly Series: Soybeans**

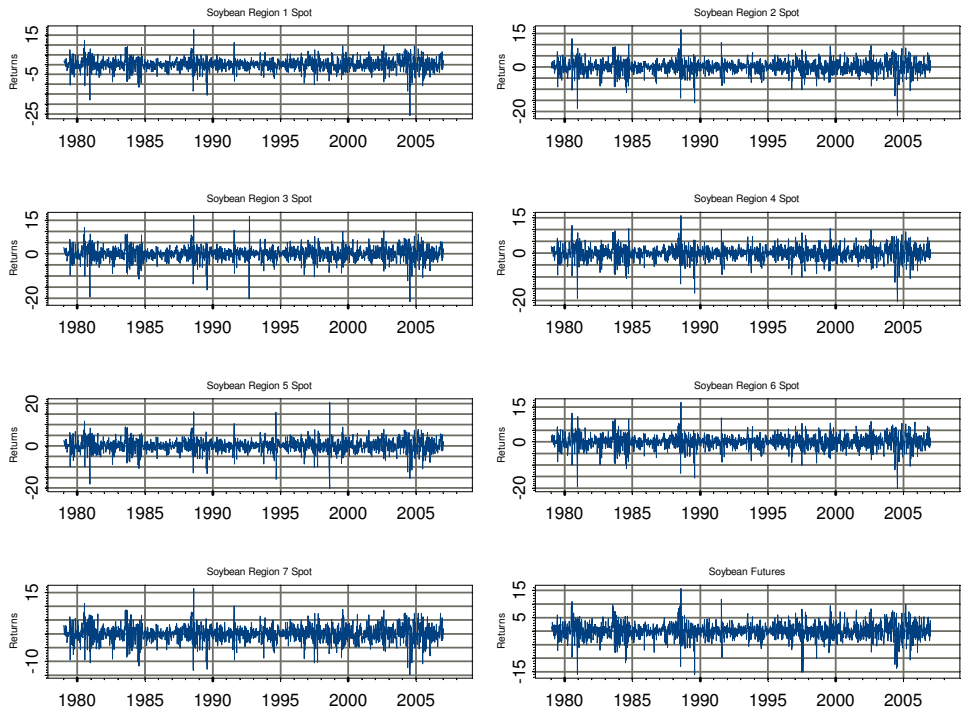

Notes: Plots show weekly % returns for each contract over the period January 1979 to December 2006. The sample size is 1461.

# Figure 2a: QQ Plots for Corn Spot and Futures Returns

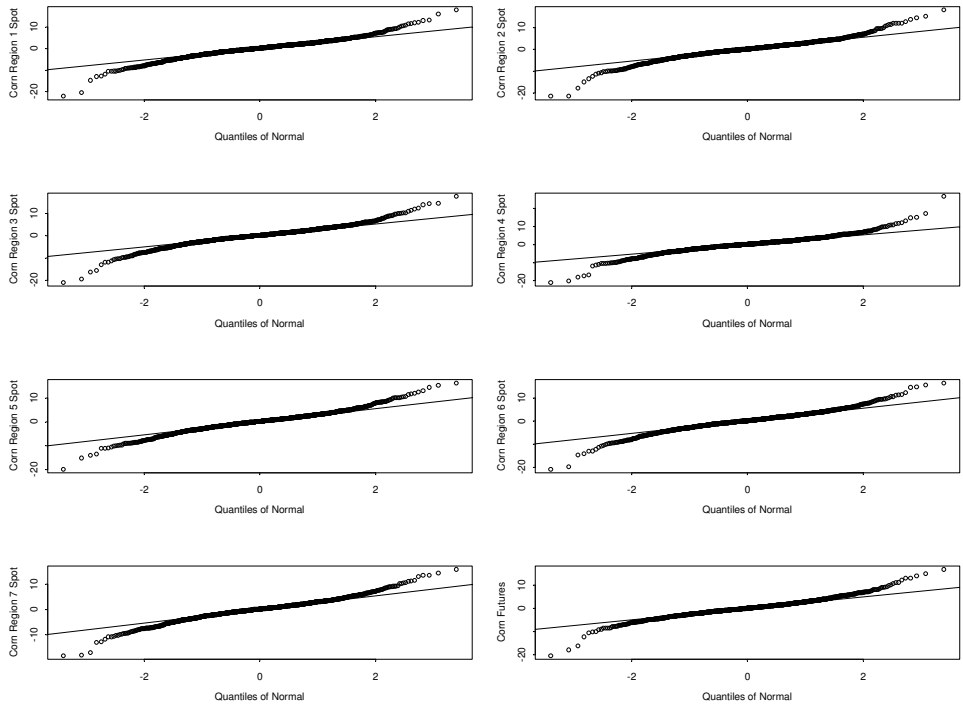

Notes: Plots show empirical quantiles of return series against those of a normal distribution. Based on 1461 weekly observations over the period January 1979 to December 2006.

**Figure 2b: QQ Plots for Soybean Spot and Futures Returns**

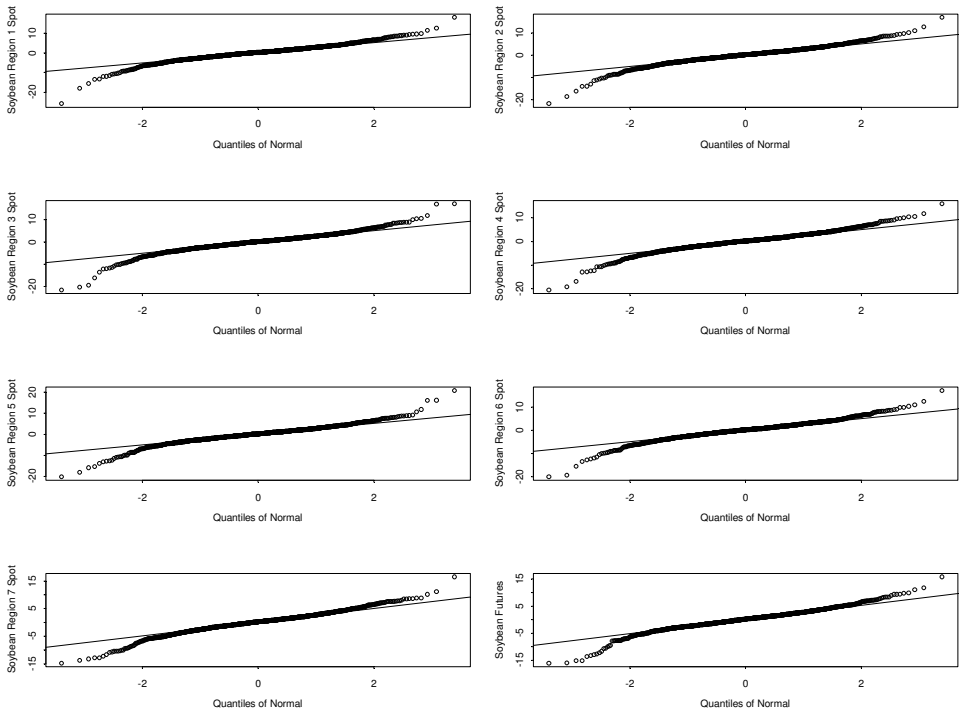

Notes: Plots show empirical quantiles of return series against those of a normal distribution. Based on 1461 weekly observations over the period January 1979 to December 2006.

**Figure 3a: Tail Index Plots as Functions of Numbers of Exceedances: Corn Spot and Futures**

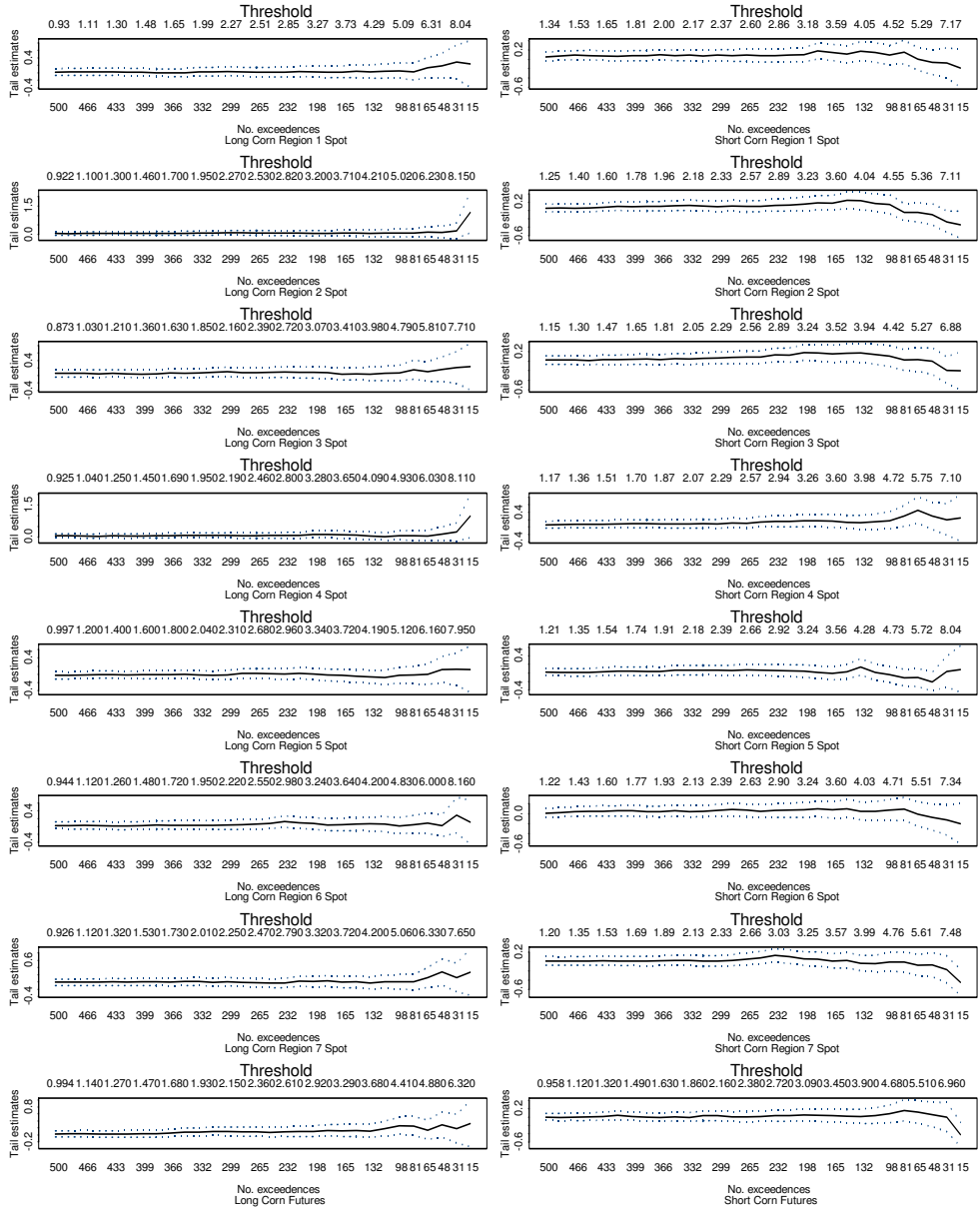

Notes: Plots show tail index ($\xi$) estimates and 95% confidence bands are presented as a function of threshold size and number of exceedences. Based on 1461 weekly observations over the period January 1979 to December 2006.

**Figure 3b: Tail Index Plots as Functions of Numbers of Exceedances: Soybean Spot and Futures**

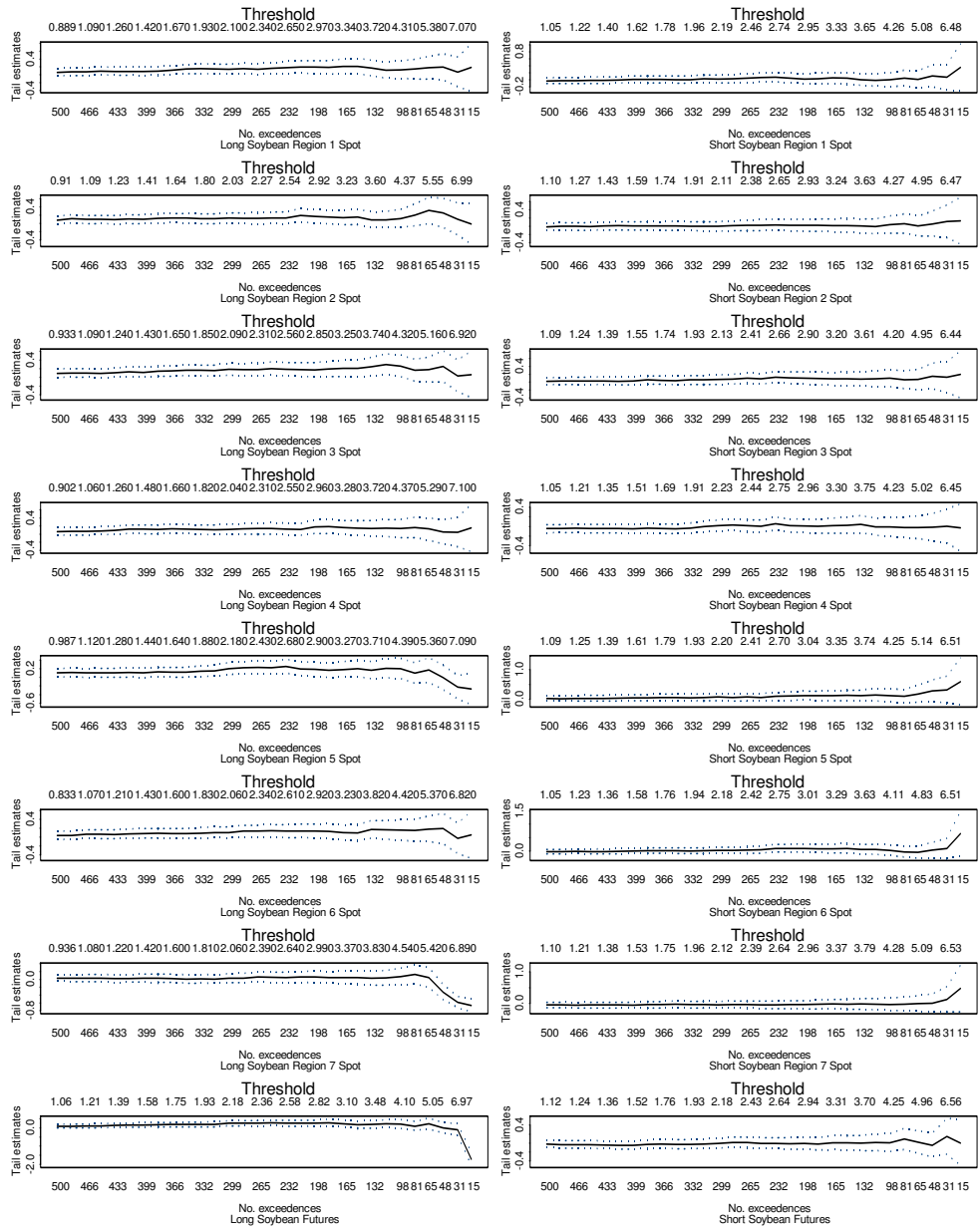

Notes: Plots show tail index ($\xi$) estimates and 95% confidence bands are presented as a function of threshold size and number of exceedances. Based on 1461 weekly observations over the period January 1979 to December 2006.

**Figure 4a: Exceedances Fitted to GPD: Corn Spot and Futures**

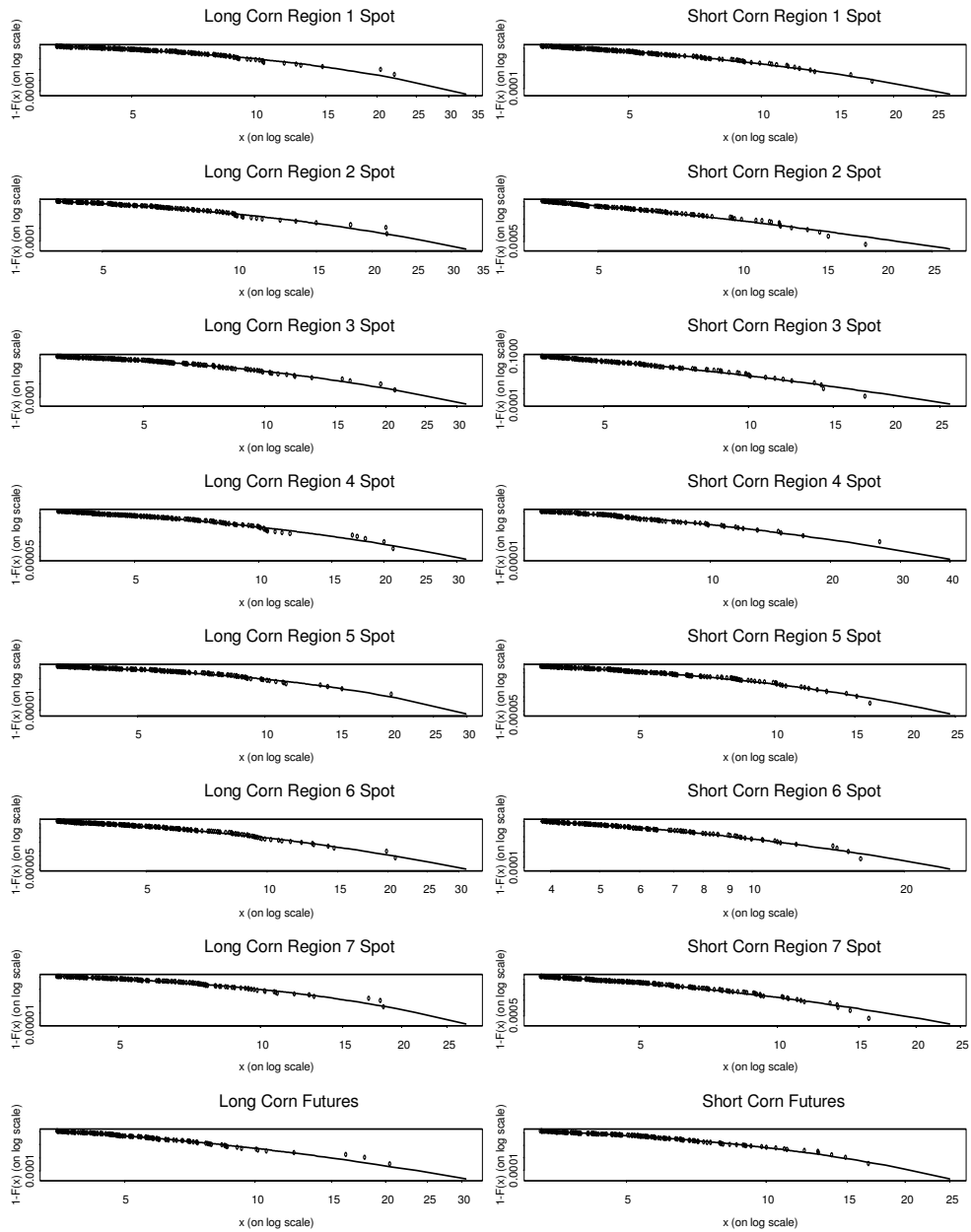

Notes: Plots show empirical exceedances against GPD-fitted exceedance curves. Based on 1461 weekly observations over the period January 1979 to December 2006.

**Figure 4b: Exceedances Fitted to GPD: Soybean Spot and Futures**

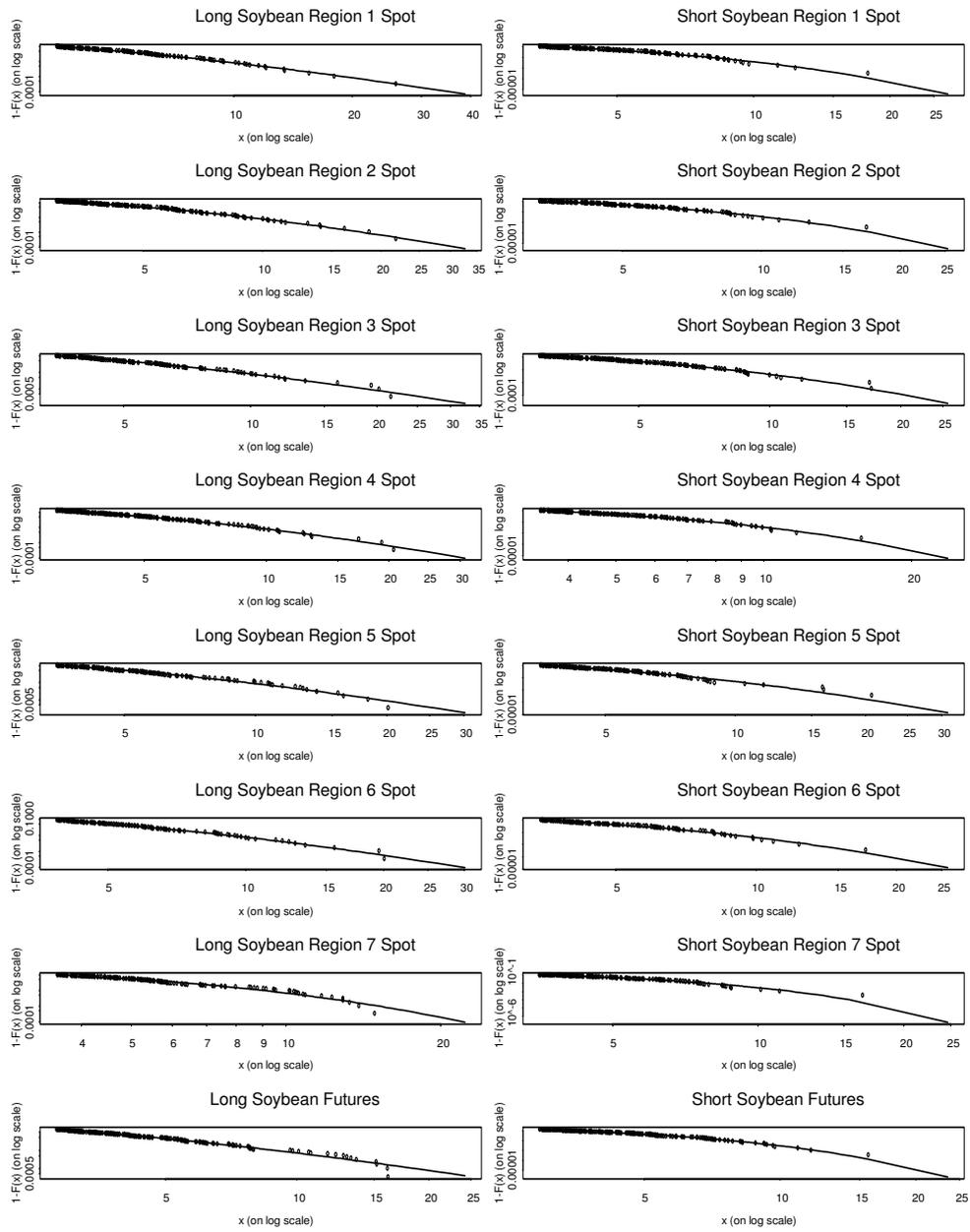

Notes: Plots show empirical exceedances against GPD-fitted exceedance curves. Based on 1461 weekly observations over the period January 1979 to December 2006.

# TABLES

### Table 1: Existing Studies of Measures of Agricultural Financial Risk

| Study | Application | Data | Estimation method |
|---|---|---|---|
| Manfredo and Leuthold (2001) | US cattle market | Weekly cash prices | Parametric methods, including RiskMetrics, and Garch and implied volatility estimates of volatility. Historical simulation |
| Odening and Mußhoff (2002) | German hog market | Weekly prices | Multivariate parametric methods including EWM A and GARCH volatility models. Historical simulation |
| Odening and Hinrichs (2003) | German hogs and farrows. Focus on cashflow-at-risk rather than VaR per se | Weekly prices. | Parametric methods with GARCH, square-root rule and Drost-Nijman formula for volatilities. Historical simulation and Generalized Extreme Value approaches |
| Pritchett et alia (2004) | Impact of alternative risk management strategies in US agriculture | Annual | Unspecified |
| Dawson and White (2005) | A typical UK arable farm | Weekly cash prices | Multivariate parametric methods, including RiskMetrics and GARCH volatility models |
| Katchova and Barry (2005) | Portfolio of Illinois farms | Annual 1995-2002 | CreditMetrics and KMV models of credit quality used to estimate default VaRs |
| Siaplay et alia (2005) | US turkey market, with the emphasis on food safety | Monthly prices and costs | Various parametric methods (including Extreme Value distribution) estimated using @Risk software |
| Wilson et alia (2005) | US bakeries | Monthly | Monte Carlo |
| Zhang et alia (2007) | Applies downside risk management techniques to investigate how US Govt. policies affect a typical farm's financial risk management | Cotton in Colquitt County GA Daily futures prices | Monte Carlo simulation, conditional kernel approach, copula methods |

**Table 2: Summary Statistics for Weekly Series**

|            | Mean   | Std Dev | Skewness | Kurtosis | JB P-value |
|------------|--------|---------|----------|----------|------------|
|            |        |         | Corn     |          |            |
| Reg 1 Spot | 0.033  | 3.495   | -0.331   | 6.557    | 0          |
| Reg 2 Spot | 0.033  | 3.554   | -0.342   | 7.022    | 0          |
| Reg 3 Spot | 0.034  | 3.402   | -0.347   | 6.942    | 0          |
| Reg 4 Spot | 0.033  | 3.585   | -0.153   | 8.540    | 0          |
| Reg 5 Spot | 0.029  | 3.512   | -0.109   | 5.491    | 0          |
| Reg 6 Spot | 0.030  | 3.497   | -0.279   | 6.362    | 0          |
| Reg 7 Spot | 0.029  | 3.485   | -0.219   | 5.698    | 0          |
| Futures    | 0.032  | 3.205   | 0.005    | 6.857    | 0          |
|            |        |         | Soybean  |          |            |
| Reg 1 Spot | -0.001 | 3.224   | -0.640   | 8.379    | 0          |
| Reg 2 Spot | 0.000  | 3.166   | -0.577   | 7.435    | 0          |
| Reg 3 Spot | -0.001 | 3.210   | -0.597   | 8.488    | 0          |
| Reg 4 Spot | -0.001 | 3.169   | -0.571   | 7.058    | 0          |
| Reg 5 Spot | -0.001 | 3.272   | -0.393   | 7.849    | 0          |
| Reg 6 Spot | 0.000  | 3.161   | -0.516   | 7.092    | 0          |
| Reg 7 Spot | -0.001 | 3.127   | -0.383   | 5.404    | 0          |
| Futures    | -0.001 | 3.100   | -0.444   | 6.359    | 0          |

Notes: Based on 1462 weekly % return observations for each of the stated series indexes over the period January 1979 through December 2006. Mean and standard deviation are in percentage form. 'JB P-value' is the P-value of the Jarque-Bera normality test.

**Table 3: Estimated Risk Measures Under the Assumption that Returns are Normal**

|  | VaR at $\alpha =$ | | | ES at $\alpha =$ | | | SRM at $\gamma$ ARA= | | |
|---|---|---|---|---|---|---|---|---|---|
|  | 0.99 | 0.995 | 0.999 | 0.99 | 0.995 | 0.999 | 20 | 100 | 200 |
|  | Corn | | | | | | | | |
| Reg 1 Spot | 8.098 | 8.970 | 10.767 | 9.282 | 10.074 | 11.735 | 6.512 | 8.788 | 9.624 |
| Reg 2 Spot | 8.235 | 9.122 | 10.950 | 9.439 | 10.245 | 11.934 | 6.621 | 8.936 | 9.786 |
| Reg 3 Spot | 7.880 | 8.729 | 10.479 | 9.033 | 9.804 | 11.421 | 6.340 | 8.556 | 9.370 |
| Reg 4 Spot | 8.307 | 9.201 | 11.046 | 9.522 | 10.335 | 12.038 | 6.678 | 9.014 | 9.871 |
| Reg 5 Spot | 8.141 | 9.017 | 10.824 | 9.331 | 10.128 | 11.796 | 6.539 | 8.827 | 9.666 |
| Reg 6 Spot | 8.105 | 8.978 | 10.777 | 9.290 | 10.083 | 11.745 | 6.512 | 8.790 | 9.626 |
| Reg 7 Spot | 8.078 | 8.948 | 10.741 | 9.259 | 10.049 | 11.705 | 6.489 | 8.759 | 9.592 |
| Futures | 7.424 | 8.224 | 9.872 | 8.510 | 9.237 | 10.760 | 5.973 | 8.061 | 8.827 |
|  | Soybeans | | | | | | | | |
| Reg 1 Spot | 7.501 | 8.306 | 9.964 | 8.594 | 9.325 | 10.857 | 5.975 | 8.075 | 8.846 |
| Reg 2 Spot | 7.365 | 8.155 | 9.784 | 8.438 | 9.156 | 10.660 | 5.869 | 7.931 | 8.688 |
| Reg 3 Spot | 7.469 | 8.269 | 9.921 | 8.556 | 9.284 | 10.809 | 5.949 | 8.040 | 8.808 |
| Reg 4 Spot | 7.373 | 8.164 | 9.794 | 8.447 | 9.166 | 10.671 | 5.873 | 7.938 | 8.695 |
| Reg 5 Spot | 7.613 | 8.429 | 10.112 | 8.722 | 9.464 | 11.018 | 6.064 | 8.196 | 8.978 |
| Reg 6 Spot | 7.354 | 8.142 | 9.768 | 8.425 | 9.141 | 10.643 | 5.859 | 7.919 | 8.674 |
| Reg 7 Spot | 7.275 | 8.055 | 9.663 | 8.334 | 9.043 | 10.529 | 5.796 | 7.833 | 8.581 |
| Futures | 7.213 | 7.986 | 9.581 | 8.263 | 8.966 | 10.439 | 5.745 | 7.765 | 8.506 |

Notes: Based on 1462 weekly % return observations for each of the stated series indexes over the period January 1979 through December 2006. Estimates of SRMs obtained using the CompEcon software of Miranda and Fackler (2002) written in MATLAB using the trapezoidal rule and $N$=1m 'slices'.

## Table 4: GPD Parameters for Weekly Series

| | Long Position | | | | | Short Position | | | | |
|---|---|---|---|---|---|---|---|---|---|---|
| | $u$ | $prob$ | $N_u$ | $\hat{\xi}$ (tail) | $\hat{\beta}$ (scale) | $u$ | $prob$ | $N_u$ | $\hat{\xi}$ (tail) | $\hat{\beta}$ (scale) |
| | | | | | **Corn** | | | | | |
| Reg 1 Spot | 3.269 | 0.862 | 201 | 0.036 | 2.445 | 3.153 | 0.863 | 200 | 0.089 | 1.978 |
| | | | | (0.068) | (0.239) | | | | (0.078) | (0.208) |
| Reg 2 Spot | 3.957 | 0.897 | 150 | 0.073 | 2.478 | 3.793 | 0.897 | 150 | 0.207 | 1.813 |
| | | | | (0.085) | (0.292) | | | | (0.113) | (0.250) |
| Reg 3 Spot | 3.052 | 0.863 | 200 | 0.084 | 2.320 | 3.697 | (0.897 | 150 | 0.167 | 1.786 |
| | | | | (0.078) | (0.243) | | | | (0.106) | (0.238) |
| Reg 4 Spot | 3.238 | 0.863 | 200 | 0.118 | 2.293 | 3.748 | 0.897 | 150 | 0.135 | 2.080 |
| | | | | (0.068) | (0.239) | | | | (0.078) | (0.208) |
| Reg 5 Spot | 3.223 | 0.856 | 210 | 0.016 | 2.357 | 3.031 | 0.849 | 220 | 0.056 | 2.165 |
| | | | | (0.073) | (0.237) | | | | (0.080) | (0.226) |
| Reg 6 Spot | 2.993 | 0.843 | 230 | 0.120 | 2.104 | 3.822 | 0.897 | 150 | 0.091 | 2.037 |
| | | | | (0.080) | (0.217) | | | | (0.098) | (0.260) |
| Reg 7 Spot | 3.685 | 0.884 | 170 | 0.012 | 2.454 | 3.048 | 0.843 | 230 | 0.130 | 1.828 |
| | | | | (0.076) | (0.264) | | | | (0.087) | (0.200) |
| Futures | 3.484 | 0.897 | 150 | 0.132 | 1.781 | 3.256 | 0.877 | 180 | 0.033 | 2.162 |
| | | | | (0.084) | (0.208) | | | | (0.078) | (0.234) |
| | | | | | **Soybean** | | | | | |
| Reg 1 Spot | 3.550 | 0.897 | 150 | 0.229 | 1.875 | 3.377 | 0.890 | 160 | 0.040 | 1.843 |
| | | | | (0.109) | (0.254) | | | | (0.082) | (0.210) |
| Reg 2 Spot | 3.008 | 0.870 | 190 | 0.177 | 1.921 | 3.308 | 0.890 | 160 | 0.022 | 1.863 |
| | | | | (0.090) | (0.221) | | | | (0.078) | (0.207) |
| Reg 3 Spot | 3.462 | 0.897 | 150 | 0.223 | 1.921 | 2.951 | 0.870 | 190 | 0.083 | 1.795 |
| | | | | (0.102) | (0.248) | | | | (0.078) | (0.191) |
| Reg 4 Spot | 3.043 | 0.870 | 190 | 0.178 | 1.903 | 3.494 | 0.897 | 150 | 0.028 | 1.801 |
| | | | | (0.109) | (0.254) | | | | (0.082) | (0.210) |
| Reg 5 Spot | 3.506 | 0.897 | 150 | 0.205 | 2.021 | 3.525 | 0.897 | 150 | 0.116 | 1.752 |
| | | | | (0.107) | (0.270) | | | | (0.083) | (0.203) |
| Reg 6 Spot | 3.870 | 0.911 | 130 | 0.179 | 1.958 | 3.433 | 0.897 | 150 | 0.075 | 1.732 |
| | | | | (0.109) | (0.272) | | | | (0.095) | (0.216) |
| Reg 7 Spot | 3.580 | 0.897 | 150 | 0.052 | 2.200 | 3.533 | 0.897 | 150 | -0.023 | 1.856 |
| | | | | (0.100) | (0.283) | | | | (0.068) | (0.197) |
| Futures | 2.821 | 0.863 | 200 | 0.252 | 1.627 | 2.934 | 0.863 | 200 | 0.000 | 1.842 |
| | | | | (0.095) | (0.191) | | | | (0.007) | (0.131) |

Notes: The Table presents estimates of the GPD parameters for long and short positions in spot and futures corn and soybean contracts. The sample size $n$ is 1462, the threshold is $u$, the probability of an observation in excess of $u$ is *prob*, the number of exceedences in excess of $u$ is $N_u$, the estimated tail parameter is $\hat{\xi}$ and the estimated scale parameter is $\hat{\beta}$. The numbers in brackets are the estimated standard errors of the parameters concerned. The thresholds $u$ are chosen as the approximate points where the QQ plots for each series change slope.

**Table 5a: GPD Values-at-Risk and Expected Shortfalls: Corn Spot and Futures Contracts**

|  | Long positions | | | Short positions | | |
|---|---|---|---|---|---|---|
|  | $\alpha=0.99$ | $\alpha=0.995$ | $\alpha=0.999$ | $\alpha=0.99$ | $\alpha=0.995$ | $\alpha=0.999$ |
|  | **Values-at-Risk** | | | | | |
| Region 1 spot | 9.989 | 11.875 | 16.440 | 8.979 | 10.764 | 15.359 |
| SE | 0.678 | 1.008 | 2.304 | 0.629 | 0.975 | 2.458 |
| Region 2 spot | 10.246 | 12.334 | 17.610 | 7.327 | 8.752 | 12.969 |
| SE | 0.741 | 1.133 | 2.772 | 0.482 | 0.819 | 2.595 |
| Region 3 spot | 9.839 | 11.902 | 17.181 | 8.779 | 10.716 | 16.178 |
| SE | 0.729 | 1.124 | 2.808 | 0.664 | 1.093 | 3.199 |
| Region 4 spot | 10.265 | 12.520 | 18.526 | 9.438 | 11.508 | 17.130 |
| SE | 0.787 | 1.247 | 3.320 | 0.718 | 1.153 | 3.170 |
| Region 5 spot | 9.640 | 11.354 | 15.409 | 9.370 | 11.151 | 15.563 |
| SE | 0.621 | 0.909 | 2.004 | 0.636 | 0.960 | 2.276 |
| Region 6 spot | 9.865 | 11.981 | 17.632 | 9.105 | 10.906 | 15.554 |
| SE | 0.738 | 1.171 | 3.131 | 0.635 | 0.985 | 2.492 |
| Region 7 spot | 9.795 | 11.554 | 15.696 | 9.106 | 11.003 | 16.127 |
| SE | 0.638 | 0.931 | 2.038 | 0.659 | 1.054 | 2.872 |
| Futures | 8.338 | 10.096 | 14.855 | 8.915 | 10.562 | 14.534 |
| SE | 0.610 | 0.978 | 2.674 | 0.593 | 0.879 | 1.998 |
|  | **Expected Shortfalls** | | | | | |
| Region 1 spot | 12.777 | 14.733 | 19.468 | 11.720 | 13.679 | 18.723 |
| SE | 0.703 | 1.045 | 2.390 | 0.691 | 1.070 | 2.698 |
| Region 2 spot | 13.414 | 15.667 | 21.359 | 9.739 | 11.537 | 16.855 |
| SE | 0.799 | 1.222 | 2.991 | 0.607 | 1.033 | 3.272 |
| Region 3 spot | 12.995 | 15.247 | 21.010 | 11.942 | 14.267 | 20.824 |
| SE | 0.796 | 1.227 | 3.066 | 0.797 | 1.313 | 3.841 |
| Region 4 spot | 13.805 | 16.362 | 23.171 | 12.731 | 15.124 | 21.623 |
| SE | 0.892 | 1.414 | 3.764 | 0.830 | 1.333 | 3.665 |
| Region 5 spot | 12.139 | 13.882 | 18.002 | 12.039 | 13.926 | 18.600 |
| SE | 0.631 | 0.923 | 2.036 | 0.673 | 1.017 | 2.410 |
| Region 6 spot | 13.193 | 15.598 | 22.019 | 11.874 | 13.856 | 18.969 |
| SE | 0.839 | 1.331 | 3.558 | 0.698 | 1.083 | 2.741 |
| Region 7 spot | 12.353 | 14.134 | 18.325 | 12.112 | 14.293 | 20.182 |
| SE | 0.646 | 0.942 | 2.063 | 0.758 | 1.212 | 3.301 |
| Futures | 11.129 | 13.154 | 18.636 | 11.344 | 13.047 | 17.155 |
| SE | 0.703 | 1.126 | 3.081 | 0.613 | 0.909 | 2.067 |

Notes: Based on 1462 weekly % return observations for each of the stated series indexes over the period January 1979 through December 2006. $\alpha$ indicates the confidence level and SE indicates the standard error of the risk measure in the box above. Standard errors are based on 5000 semi-parametric bootstrap resamples.

**Table 5b: GPD Values-at-Risk and Expected Shortfalls: Soybean Spot and Futures Contracts**

|  | Long positions | | | Short positions | | |
|---|---|---|---|---|---|---|
|  | $\alpha=0.99$ | $\alpha=0.995$ | $\alpha=0.999$ | $\alpha=0.99$ | $\alpha=0.995$ | $\alpha=0.999$ |
| | **Values-at-Risk** | | | | | |
| Region 1 spot | 9.317 | 11.717 | 19.006 | 8.005 | 9.430 | 12.897 |
| SE | 0.805 | 1.393 | 4.611 | 0.512 | 0.763 | 1.757 |
| Region 2 spot | 9.243 | 11.474 | 17.841 | 7.885 | 9.257 | 12.523 |
| SE | 0.762 | 1.265 | 3.775 | 0.496 | 0.729 | 1.624 |
| Region 3 spot | 9.326 | 11.746 | 19.042 | 8.081 | 9.666 | 13.716 |
| SE | 0.813 | 1.401 | 4.580 | 0.560 | 0.863 | 2.152 |
| Region 4 spot | 9.228 | 11.444 | 17.778 | 7.827 | 9.172 | 12.399 |
| SE | 0.757 | 1.257 | 3.759 | 0.485 | 0.717 | 1.615 |
| Region 5 spot | 9.537 | 11.963 | 19.122 | 8.208 | 9.865 | 14.266 |
| SE | 0.820 | 1.393 | 4.394 | 0.578 | 0.915 | 2.428 |
| Region 6 spot | 9.106 | 11.243 | 17.356 | 7.839 | 9.306 | 13.023 |
| SE | 0.729 | 1.213 | 3.633 | 0.520 | 0.797 | 1.957 |
| Region 7 spot | 9.025 | 10.778 | 15.099 | 7.741 | 8.950 | 11.686 |
| SE | 0.626 | 0.943 | 2.219 | 0.445 | 0.631 | 1.298 |
| Futures | 8.847 | 11.229 | 18.663 | 7.753 | 9.029 | 11.994 |
| SE | 0.792 | 1.397 | 4.846 | 0.465 | 0.672 | 1.441 |
| | **Expected Shortfalls** | | | | | |
| Region 1 spot | 13.462 | 16.575 | 26.029 | 10.118 | 11.602 | 15.213 |
| SE | 1.044 | 1.807 | 5.981 | 0.533 | 0.795 | 1.830 |
| Region 2 spot | 12.918 | 15.629 | 23.365 | 9.893 | 11.296 | 14.635 |
| SE | 0.926 | 1.537 | 4.587 | 0.507 | 0.745 | 1.661 |
| Region 3 spot | 13.481 | 16.596 | 25.985 | 10.503 | 12.231 | 16.648 |
| SE | -1.046 | 1.803 | 5.895 | 0.611 | 0.941 | 2.347 |
| Region 4 spot | 12.883 | 15.579 | 23.284 | 9.805 | 11.189 | 14.508 |
| SE | 0.920 | 1.530 | 4.574 | 0.499 | 0.737 | 1.661 |
| Region 5 spot | 13.634 | 16.685 | 25.690 | 10.805 | 12.679 | 17.658 |
| SE | 1.032 | 1.752 | 5.526 | 0.654 | 1.035 | 2.746 |
| Region 6 spot | 12.632 | 15.235 | 22.681 | 10.069 | 11.655 | 15.673 |
| SE | 0.888 | 1.477 | 4.426 | 0.562 | 0.861 | 2.116 |
| Region 7 spot | 11.645 | 13.493 | 18.052 | 9.460 | 10.643 | 13.317 |
| SE | 0.661 | 0.994 | 2.341 | 0.435 | 0.617 | 1.269 |
| Futures | 13.052 | 16.237 | 26.176 | 9.595 | 10.872 | 13.836 |
| SE | 1.059 | 1.867 | 6.478 | 0.465 | 0.672 | 1.441 |

Notes: Based on 1462 weekly % return observations for each of the stated series indexes over the period January 1979 through December 2006. $\alpha$ indicates the confidence level and SE indicates the standard error of the risk measure in the box above. Standard errors are based on 5000 semi-parametric bootstrap resamples.

**Table 6a: Standardised 90% Confidence Intervals for Values-at-Risk and Expected Shortfalls: Corn Spot and Futures Contracts**

| contract | Long positions | | | | | | Short positions | | | | | |
|---|---|---|---|---|---|---|---|---|---|---|---|---|
| | $\alpha=0.99$ | | $\alpha=0.995$ | | $\alpha=0.999$ | | $\alpha=0.99$ | | $\alpha=0.995$ | | $\alpha=0.999$ | |
| | LB | UB | LB | UB | LB | UB | LB | UB | LB | UB | LB | UB |
| **Values-at-Risk** | | | | | | | | | | | | |
| region 1 spot | 0.893 | 1.117 | 0.871 | 1.148 | 0.800 | 1.255 | 0.890 | 1.121 | 0.864 | 1.159 | 0.780 | 1.293 |
| region 2 spot | 0.886 | 1.125 | 0.861 | 1.161 | 0.781 | 1.288 | 0.899 | 1.115 | 0.864 | 1.166 | 0.748 | 1.366 |
| region 3 spot | 0.884 | 1.128 | 0.858 | 1.165 | 0.774 | 1.299 | 0.883 | 1.132 | 0.850 | 1.180 | 0.743 | 1.363 |
| region 4 spot | 0.880 | 1.133 | 0.851 | 1.175 | 0.758 | 1.329 | 0.882 | 1.132 | 0.851 | 1.176 | 0.754 | 1.340 |
| region 5 spot | 0.898 | 1.110 | 0.877 | 1.139 | 0.812 | 1.236 | 0.893 | 1.117 | 0.870 | 1.150 | 0.794 | 1.267 |
| region 6 spot | 0.883 | 1.130 | 0.854 | 1.172 | 0.761 | 1.326 | 0.891 | 1.121 | 0.864 | 1.158 | 0.780 | 1.293 |
| region 7 spot | 0.897 | 1.112 | 0.877 | 1.140 | 0.812 | 1.236 | 0.887 | 1.126 | 0.858 | 1.168 | 0.762 | 1.327 |
| futures | 0.886 | 1.127 | 0.856 | 1.170 | 0.760 | 1.330 | 0.895 | 1.114 | 0.873 | 1.145 | 0.804 | 1.250 |
| **Expected Shortfalls** | | | | | | | | | | | | |
| region 1 spot | 0.913 | 1.094 | 0.892 | 1.123 | 0.825 | 1.223 | 0.908 | 1.102 | 0.882 | 1.137 | 0.802 | 1.264 |
| region 2 spot | 0.907 | 1.103 | 0.882 | 1.136 | 0.805 | 1.256 | 0.904 | 1.109 | 0.870 | 1.159 | 0.756 | 1.355 |
| region 3 spot | 0.904 | 1.106 | 0.879 | 1.141 | 0.799 | 1.267 | 0.897 | 1.116 | 0.865 | 1.162 | 0.760 | 1.338 |
| region 4 spot | 0.899 | 1.112 | 0.871 | 1.152 | 0.781 | 1.298 | 0.899 | 1.113 | 0.869 | 1.155 | 0.774 | 1.311 |
| region 5 spot | 0.918 | 1.089 | 0.898 | 1.115 | 0.837 | 1.205 | 0.912 | 1.096 | 0.889 | 1.127 | 0.818 | 1.236 |
| region 6 spot | 0.901 | 1.110 | 0.873 | 1.150 | 0.783 | 1.296 | 0.908 | 1.101 | 0.883 | 1.137 | 0.802 | 1.265 |
| region 7 spot | 0.917 | 1.089 | 0.898 | 1.116 | 0.837 | 1.204 | 0.903 | 1.108 | 0.874 | 1.149 | 0.782 | 1.300 |
| futures | 0.902 | 1.110 | 0.873 | 1.151 | 0.780 | 1.303 | 0.915 | 1.093 | 0.894 | 1.121 | 0.828 | 1.219 |

Notes: Based on 1462 weekly % return observations for each of the stated series indexes over the period January 1979 through December 2006, and based on 5000 semi-parametric bootstrap resamples. $\alpha$ indicates the confidence level, and LB and UB refer to the lower and upper bounds of the 90% confidence interval divided by the estimated mean of the risk measure concerned.

**Table 6b: Standardised 90% Confidence Intervals for Values-at-Risk and Expected Shortfalls: Soybean Spot and Futures Contracts**

| | Long positions | | | | | | Short positions | | | | | |
|---|---|---|---|---|---|---|---|---|---|---|---|---|
| | $\alpha=0.99$ | | $\alpha=0.995$ | | $\alpha=0.999$ | | $\alpha=0.99$ | | $\alpha=0.995$ | | $\alpha=0.999$ | |
| contract | LB | UB | LB | UB | LB | UB | LB | UB | LB | UB | LB | UB |
| | **Values-at-Risk** | | | | | | | | | | | |
| region 1 spot | 0.867 | 1.152 | 0.828 | 1.211 | 0.701 | 1.443 | 0.899 | 1.110 | 0.877 | 1.141 | 0.806 | 1.248 |
| region 2 spot | 0.872 | 1.144 | 0.838 | 1.195 | 0.727 | 1.388 | 0.901 | 1.108 | 0.880 | 1.137 | 0.814 | 1.236 |
| region 3 spot | 0.866 | 1.153 | 0.827 | 1.211 | 0.702 | 1.439 | 0.891 | 1.120 | 0.866 | 1.156 | 0.783 | 1.287 |
| region 4 spot | 0.873 | 1.143 | 0.839 | 1.194 | 0.728 | 1.388 | 0.902 | 1.106 | 0.881 | 1.136 | 0.814 | 1.237 |
| region 5 spot | 0.867 | 1.151 | 0.830 | 1.206 | 0.711 | 1.421 | 0.890 | 1.122 | 0.862 | 1.163 | 0.770 | 1.312 |
| region 6 spot | 0.876 | 1.140 | 0.842 | 1.191 | 0.731 | 1.384 | 0.896 | 1.114 | 0.871 | 1.150 | 0.791 | 1.275 |
| region 7 spot | 0.891 | 1.120 | 0.867 | 1.152 | 0.793 | 1.268 | 0.909 | 1.098 | 0.891 | 1.122 | 0.836 | 1.200 |
| futures | 0.863 | 1.158 | 0.821 | 1.221 | 0.686 | 1.472 | 0.905 | 1.103 | 0.886 | 1.129 | 0.825 | 1.217 |
| | **Expected Shortfalls** | | | | | | | | | | | |
| region 1 spot | 0.881 | 1.136 | 0.842 | 1.193 | 0.716 | 1.420 | 0.917 | 1.090 | 0.896 | 1.119 | 0.829 | 1.219 |
| region 2 spot | 0.889 | 1.125 | 0.856 | 1.174 | 0.747 | 1.360 | 0.919 | 1.088 | 0.899 | 1.114 | 0.837 | 1.206 |
| region 3 spot | 0.881 | 1.136 | 0.843 | 1.193 | 0.719 | 1.415 | 0.909 | 1.100 | 0.884 | 1.135 | 0.805 | 1.258 |
| region 4 spot | 0.890 | 1.125 | 0.856 | 1.173 | 0.747 | 1.360 | 0.920 | 1.087 | 0.899 | 1.114 | 0.836 | 1.208 |
| region 5 spot | 0.884 | 1.132 | 0.847 | 1.186 | 0.729 | 1.394 | 0.906 | 1.105 | 0.878 | 1.143 | 0.790 | 1.285 |
| region 6 spot | 0.891 | 1.123 | 0.858 | 1.171 | 0.749 | 1.358 | 0.913 | 1.096 | 0.889 | 1.129 | 0.813 | 1.247 |
| region 7 spot | 0.911 | 1.097 | 0.888 | 1.128 | 0.817 | 1.236 | 0.927 | 1.078 | 0.910 | 1.100 | 0.859 | 1.172 |
| futures | 0.876 | 1.143 | 0.835 | 1.204 | 0.701 | 1.450 | 0.923 | 1.083 | 0.905 | 1.107 | 0.848 | 1.188 |

Notes: Based on 1462 weekly % return observations for each of the stated series indexes over the period January 1979 through December 2006, and based on 5000 semi-parametric bootstrap resamples. $\alpha$ indicates the confidence level, and LB and UB refer to the lower and upper bounds of the 90% confidence interval divided by the estimated mean of the risk measure concerned.

**Table 7: Spectral Risk Measure Estimates and % Errors**

| Numerical Integration Method | Spectral Risk Measure (SRM) Estimates | | | | | |
|---|---|---|---|---|---|---|
| | $N=1000$ | $N=10,000$ | $N=100,000$ | $N=1m$ | $N=10m$ | $N=20m$ |
| Trapezoidal rule | 8.926 | 10.451 | 10.693 | 10.728 | 10.733 | 10.733 |
| Simpson's rule | 8.894 | 10.448 | 10.693 | 10.728 | 10.733 | 10.733 |
| Niederreiter QMC | 9.154 | 10.340 | 10.668 | 10.725 | 10.733 | 10.733 |
| Weyl QMC | 9.154 | 10.340 | 10.668 | 10.725 | 10.733 | 10.733 |
| | % errors in SRM estimates | | | | | |
| Trapezoidal rule | $N=1000$ | $N=10,000$ | $N=100,000$ | $N=1m$ | $N=10m$ | |
| Simpson's rule | -16.835 | -2.628 | -0.372 | -0.048 | -0.003 | NA |
| Niederreiter QMC | -17.131 | -2.658 | -0.376 | -0.048 | -0.003 | NA |
| Weyl QMC | -14.712 | -3.666 | -0.610 | -0.075 | -0.005 | NA |
| Trapezoidal rule | -14.712 | -3.666 | -0.610 | -0.075 | -0.005 | NA |

Notes: Based on the mean parameters from Table 1 (i.e., $\beta=1.98$, $\xi=0.1042$, threshold = 3.3701 and $N_u=173.7813$) and $R$ (coefficient of absolute risk aversion) =100, where $N$ is the number of slices in the numerical integration. Errors are assessed against a 'true' value obtained using $N=20m$. Calculations carried out using the CompEcon software of Miranda and Fackler (2002) written in MATLAB using the trapezoidal rule.

## Table 8a: Spectral Risk Measures and Associated Precision Statistics for Corn Spot and Futures

| | Long Position | | | | | | Short position | | | | | |
|---|---|---|---|---|---|---|---|---|---|---|---|---|
| | $R=20$ | | $R=100$ | | $R=200$ | | $R=20$ | | $R=100$ | | $R=200$ | |
| | UB | LB | UB | LB | UB | UB | LB | UB | LB | UB | LB | UB |
| Region 1 | 7.344 | | 11.635 | | 13.558 | | 6.691 | | 10.655 | | 12.542 | |
| SE | 0.435 | | 1.423 | | 2.289 | | 0.398 | | 1.330 | | 2.166 | |
| CI | 0.903 | 1.097 | 0.808 | 1.205 | 0.737 | 1.288 | 0.903 | 1.097 | 0.806 | 1.210 | 0.733 | 1.297 |
| Region 2 | 7.494 | | 12.163 | | 14.346 | | 5.847 | | 8.910 | | 10.574 | |
| SE | 0.454 | | 1.520 | | 2.474 | | 0.338 | | 1.151 | | 1.918 | |
| CI | 0.901 | 1.099 | 0.805 | 1.209 | 0.733 | 1.296 | 0.906 | 1.095 | 0.800 | 1.218 | 0.724 | 1.317 |
| Region 3 | 7.172 | | 11.762 | | 13.935 | | 6.617 | | 10.809 | | 12.987 | |
| SE | 0.438 | | 1.482 | | 2.422 | | 0.406 | | 1.412 | | 2.356 | |
| CI | 0.900 | 1.101 | 0.804 | 1.212 | 0.732 | 1.299 | 0.901 | 1.102 | 0.797 | 1.221 | 0.722 | 1.314 |
| Region 4 | 7.516 | | 12.470 | | 14.906 | | 6.990 | | 11.510 | | 13.777 | |
| SE | 0.465 | | 1.600 | | 2.642 | | 0.430 | | 1.484 | | 2.459 | |
| CI | 0.900 | 1.102 | 0.800 | 1.216 | 0.729 | 1.306 | 0.900 | 1.102 | 0.799 | 1.217 | 0.727 | 1.308 |
| Region 5 | 7.153 | | 11.095 | | 12.822 | | 6.954 | | 10.967 | | 12.807 | |
| SE | 0.416 | | 1.339 | | 2.138 | | 0.410 | | 1.348 | | 2.176 | |
| CI | 0.904 | 1.094 | 0.811 | 1.201 | 0.739 | 1.285 | 0.903 | 1.096 | 0.808 | 1.206 | 0.736 | 1.290 |
| Region 6 | 7.293 | | 11.940 | | 14.230 | | 6.803 | | 10.800 | | 12.707 | |
| SE | 0.446 | | 1.526 | | 2.515 | | 0.404 | | 1.347 | | 2.195 | |
| CI | 0.901 | 1.101 | 0.801 | 1.216 | 0.729 | 1.305 | 0.903 | 1.097 | 0.806 | 1.210 | 0.733 | 1.297 |
| Region 7 | 7.227 | | 11.281 | | 13.049 | | 6.842 | | 10.992 | | 13.061 | |
| SE | 0.422 | | 1.363 | | 2.176 | | 0.412 | | 1.400 | | 2.307 | |
| CI | 0.904 | 1.095 | 0.811 | 1.201 | 0.739 | 1.285 | 0.902 | 1.099 | 0.802 | 1.215 | 0.730 | 1.305 |
| Futures | 6.248 | | 10.091 | | 12.011 | | 6.593 | | 10.346 | | 12.022 | |
| SE | 0.378 | | 1.289 | | 2.128 | | 0.387 | | 1.260 | | 2.023 | |
| CI | 0.901 | 1.100 | 0.801 | 1.216 | 0.729 | 1.306 | 0.903 | 1.096 | 0.809 | 1.203 | 0.737 | 1.287 |

Notes: Based on 1462 weekly % return observations for each of the stated series indexes over the period January 1979 through December 2006, and based on 5000 semi-parametric bootstrap resamples. $R$ is the coefficient of absolute risk aversion, SE indicates the standard error, CI indicates the standardised 90% confidence interval, and LB and UB refer to its bounds. Calculations carried out using the CompEcon software of Miranda and Fackler (2002) written in MATLAB using the trapezoidal rule and $N$=1m 'slices'.

## Table 8b: Spectral Risk Measures and Associated Precision Statistics for Soybean Spot and Futures

| | Long Position | | | | | | Short position | | | | | |
|---|---|---|---|---|---|---|---|---|---|---|---|---|
| | $R=20$ | | $R=100$ | | $R=200$ | | $R=20$ | | $R=100$ | | $R=200$ | |
| | UB | LB | UB | LB | UB | UB | LB | UB | LB | UB | LB | UB |
| Region 1 | 6.928 | | 12.072 | | 14.935 | | 6.019 | | 9.256 | | 10.713 | |
| SE | 0.458 | | 1.691 | | 2.904 | | 0.347 | | 1.122 | | 1.797 | |
| CI | 0.896 | 1.109 | 0.785 | 1.238 | 0.706 | 1.339 | 0.905 | 1.094 | 0.810 | 1.202 | 0.738 | 1.287 |
| Region 2 | 6.796 | | 11.615 | | 14.147 | | 5.916 | | 9.059 | | 10.446 | |
| SE | 0.436 | | 1.556 | | 2.622 | | 0.340 | | 1.090 | | 1.740 | |
| CI | 0.898 | 1.106 | 0.792 | 1.227 | 0.717 | 1.324 | 0.905 | 1.093 | 0.811 | 1.200 | 0.740 | 1.284 |
| Region 3 | 6.889 | | 12.079 | | 14.947 | | 6.029 | | 9.556 | | 11.224 | |
| SE | 0.458 | | 1.689 | | 2.897 | | 0.358 | | 1.188 | | 1.931 | |
| CI | 0.895 | 1.109 | 0.785 | 1.238 | 0.706 | 1.337 | 0.903 | 1.097 | 0.806 | 1.208 | 0.734 | 1.295 |
| Region 4 | 6.802 | | 11.588 | | 14.106 | | 5.914 | | 8.988 | | 10.353 | |
| SE | 0.435 | | 1.551 | | 2.614 | | 0.338 | | 1.081 | | 1.724 | |
| CI | 0.898 | 1.106 | 0.792 | 1.227 | 0.717 | 1.324 | 0.906 | 1.093 | 0.811 | 1.200 | 0.740 | 1.284 |
| Region 5 | 7.010 | | 12.224 | | 15.049 | | 6.182 | | 9.823 | | 11.610 | |
| SE | 0.461 | | 1.682 | | 2.868 | | 0.368 | | 1.238 | | 2.029 | |
| CI | 0.896 | 1.109 | 0.787 | 1.233 | 0.711 | 1.334 | 0.903 | 1.098 | 0.803 | 1.213 | 0.732 | 1.301 |
| Region 6 | 6.771 | | 11.384 | | 13.814 | | 5.912 | | 9.190 | | 10.726 | |
| SE | 0.428 | | 1.516 | | 2.550 | | 0.345 | | 1.131 | | 1.829 | |
| CI | 0.899 | 1.105 | 0.793 | 1.226 | 0.718 | 1.322 | 0.904 | 1.095 | 0.808 | 1.207 | 0.735 | 1.291 |
| Region 7 | 6.632 | | 10.588 | | 12.394 | | 5.869 | | 8.715 | | 9.906 | |
| SE | 0.395 | | 1.305 | | 2.110 | | 0.329 | | 1.026 | | 1.615 | |
| CI | 0.902 | 1.098 | 0.807 | 1.207 | 0.735 | 1.290 | 0.907 | 1.090 | 0.815 | 1.196 | 0.745 | 1.278 |
| Futures | 6.579 | | 11.677 | | 14.586 | | 5.851 | | 8.813 | | 10.087 | |
| SE | 0.447 | | 1.682 | | 2.919 | | 0.332 | | 1.048 | | 1.661 | |
| CI | 0.893 | 1.113 | 0.781 | 1.245 | 0.700 | 1.346 | 0.907 | 1.091 | 0.813 | 1.198 | 0.742 | 1.281 |

Notes: Based on 1462 weekly % return observations for each of the stated series indexes over the period January 1979 through December 2006, and based on 5000 semi-parametric bootstrap resamples. $R$ is the coefficient of absolute risk aversion, SE indicates the standard error, CI indicates the standardised 90% confidence interval, and LB and UB refer to its bounds. Calculations carried out using the CompEcon software of Miranda and Fackler (2002) written in MATLAB written in MATLAB using the trapezoidal and $N$=1m 'slices'.